%% file: main.tex
\documentclass[10pt,sigconf,letterpaper,screen,nonacm]{acmart}

\AtBeginDocument{%
  }

\newif\ifEditMode

\EditModetrue
\usepackage{aliases}
\usepackage{preamble}

\begin{document}

\title[]{The Writing is on the Wall: \\Analyzing the Boom of Inscriptions and Its Impact on Rollup Performance and Cost Efficiency}
\titlenote{This research article is a preliminary work of scholarship and reflects the authors’ own views and opinions. It does not necessarily reflect the views or opinions of any other person or organization, including the authors’ employer. Readers should not rely on this article for making strategic or commercial decisions, and the authors are not responsible for any losses that may result from such use.}

% %% Leaving e-mails out due to spam!

\author[K. Gogol]{Krzysztof Gogol}
\authornote{Both authors contributed equally to the paper.}
  % \email{gogol@ifi.uzh.ch}
 \affiliation{%
\institution{Matter Labs}
 \institution{University of Zurich}
 \country{ }
 }

  \author[J. Messias]{Johnnatan Messias}
\authornotemark[2]
  % \email{jmp@matterlabs.dev}
 \affiliation{%
   \institution{Matter Labs}
   \country{ }
 }

 \author[M. Silva]{Maria Inês Silva}
  % \email{msi@matterlabs.dev}
 \affiliation{%
   \institution{Matter Labs}
    \country{}
 }

 \author[B. Livshits]{Benjamin Livshits}
  % \email{bl@matterlabs.dev}
 \affiliation{%
   \institution{Matter Labs}
   \institution{Imperial College London}
     \country{}
 }

\renewcommand{\point}[1]{\par\smallskip\noindent\textbf{#1.} }

\begin{abstract}
\input{sections/abstract.tex}

\end{abstract}

\maketitle

%------------------------------------------------------------------------------
\input{sections/introduction}
\input{sections/related_work.tex}
\input{sections/background.tex}
\input{sections/dataset.tex}

\input{sections/inscriptions-characterization.tex}
\input{sections/discussion.tex}

\input{sections/conclusion.tex}
\input{sections/acknowledgments.tex}

%------------------------------------------------------------------------------

%\newpage
\bibliographystyle{ACM-Reference-Format}
\bibliography{main}

%\appendix
%\input{sections/appendix}

\end{document}

%% file: sections/abstract.tex
\begin{comment}
In the late surge of inscriptions in 2023, rollups underwent rigorous scalability and stress testing, resulting in a significant increase in transactions per second (TPS). This surge caused some rollups to experience approximately an hour of downtime, leading to delays in various transaction types, including exchanges, token purchases, and sales. Consequently, users raised concerns about transaction finality time.

Despite the attention given to rollups, there is limited knowledge about how these inscriptions impacted them, even from the users' perspective. To address this gap, we conducted a comprehensive data-driven analysis to evaluate the effects of inscriptions on rollups and whether users incurred financial losses.

Our research aims to shed light on the benefits and consequences of inscriptions during and after the boom. We believe that analyzing the chain activity generated by these inscriptions is crucial for assessing protocol and user safety, particularly concerning transaction finality time and high TPS. Through this investigation, we strive to provide valuable insights into the impact of inscriptions on rollups and their users, contributing to the optimization of their performance for the benefit of the community, especially if inscriptions become a lasting phenomenon.
\end{comment}

Late 2023 witnessed significant user activity on EVM chains, resulting in a surge in transaction activity and putting many rollups into the first live test. While some rollups performed well, some others experienced downtime during this period, affecting transaction finality time and gas fees. To address the lack of empirical research on rollups, we perform the first study during a heightened activity during the late~2023 transaction boom, as attributed to inscriptions~---~a novel technique that enables NFT and ERC-20 token creation on Bitcoin and other blockchains. 

We observe that minting inscription-based meme tokens on zkSync Era allows for trading at a fraction of the costs, compared to the Bitcoin or Ethereum networks.
We also found that the increased transaction activity, over~99\% attributed to the minting of new inscription tokens, positively affected other users of zkSync Era, resulting in lowered gas fees. Unlike L1 blockchains, ZK rollups may experience \emph{lower gas fees} with increased transaction volume. 
Lastly, the introduction of \emph{blobs}~---~a form of temporary data storage~---~decreased the gas costs of Ethereum rollups, but also raised a number of questions about the security of inscription-based tokens.

%% file: sections/introduction.tex
\section{Introduction}
\label{sec:introduction}

% ====== Intro to blockchains focussing on Ethereum and Bitcoin to give IMC reviewers a context ======
Since the introduction of Bitcoin~\cite{Nakamoto-WhitePaper2008} and later Ethereum~\cite{Wood@Ethereum}, blockchains have witnessed increased adoption, predominantly driven by their decentralization principles~\cite{angelis2019blockchain,prewett2020blockchain}. This surge has led to the development of a diverse array of decentralized applications (DApps), including decentralized exchanges (DEXes)~\cite{Daian@S&P20,adams2021uniswap,leshner2019compound}, lending protocols~\cite{Qin@FC21,Perez@FC21}, non-fungible tokens (NFTs) for art~\cite{NFTs}, and other applications such as supply chain management~\cite{Provenance-WebArticle2015,Hackett-WebArticle2017}, and decentralized governance~\cite{leshner2019compound, adams2021uniswap,Governance@MakerDAO,messias2023understanding}.

% ====== Transaction boom at the end of 2023 ======
As in all distributed systems, each blockchain can optimize only two out of three possible factors: security, decentralization, and scalability~\cite{werth2023review}. This is also known as the \emph{blockchain trilemma}, a term coined by V. Buterin~\cite{werth2023review}. Designers of the most popular blockchains like Ethereum, the first programmable blockchain network, optimized for decentralization and security, sacrificing scalability. Therefore, many new layer-1 (L1) and layer-2 (L2), or off-chain solution that relies on the L1 blockchain for reaching consensus, scaling solutions attempt to solve the trilemma, leading to the emergence of other EVM-compatible chains focusing on scalability without sacrificing the other two property. This is achieved by allowing transactions to be executed off-chain and the results (or state transition) are stored on the L1 blockchain (e.g., Ethereum).

However, many of these L2 scaling solutions were not yet fully stress-tested until late 2023, when a spike in transactions occurred: most L2 Ethereum Virtual Machine (EVM) compatible chains experienced their all-year heights of transaction amounts in November and December 2023. This sudden transaction spike was caused by a boom in what's called \emph{inscriptions}~\cite{2024DuneEVMInsctiptions} 

% ====== Intro to ordinals and inscriptions ======
These inscriptions, first introduced in March 2023 by an anonymous developer known as Domo~\cite{Domo@BRC}, are the response to the absence of smart contracts, i.e., programs that run atop of a blockchain, in the Bitcoin blockchain. These inscriptions allow for the creation of NFT- and ERC-20-style tokens in the Bitcoin blockchain that is not smart-contract enabled, leading to a notable surge in the number of transactions in mid-2023 due to a \stress{fear-of-missing-out (FOMO) effect}, subsequently causing an increase in transaction fees for all users~\cite{wang2023brc, wang2023understanding}. Thereafter, the concept of inscriptions was extended to other blockchains, however, these blockchains now are smart-contract enabled and EVM-compatible such as zkSync Era, Binance Smart Chain (BSC), Polygon, Arbitrum, Optimism, and Base.
This development triggered a significant increase in transaction-per-second (TPS) on these chains, reaching yearly transaction highs in December~2023, and inscription transactions representing over~\num{80}\% of all network transactions~\cite{2024DuneEVMInsctiptions}.

During the inscription boom at the break of~2023 and~2024, Zero Knowledge (KP) rollups~---~a type of layer-2 blockchain~---~witnessed notably lower gas costs compared to Ethereum, heightening the interest of users, as the cost of participating in the inscription-boom became considerably more affordable when compared to Bitcoin and Ethereum. Nevertheless, some roll-ups experienced downtime of approximately~\num{78} minutes~\cite{Tom-Arbitrum@Cointelegraph}, leading to users' concerns about their transactions' finality.

%\TODO{Briefly explains inscriptions here: What are they, how they work, DA, blobs, ...}

On March~13, 2024, Ethereum underwent the Dencun upgrade \cite{2024EthereumRoadmap}, which introduced \stress{blobs} as temporary data storage. Blobs decreased the gas costs at Ethereum rollups but also raised concerns about the security of inscription-based tokens. After the 18-day period, blobs disappear from the Ethereum blockchain, as they are designed to support the state verification for rollups. Consequently, inscription-related data might remain only available off-chain in the indexer of their creator if the rollup decides not to store all transaction data.

%\TODO{Add more findings here..}

%We found that minting inscription-based meme tokens on zkSync Era allows for trading at a fraction of the costs compared to the Bitcoin or Ethereum networks. %However, trading activity remains limited, with users purchasing one of the two fully minted meme tokens: \stress{bgnt} and \stress{sync} for prices in the range of~\$0.1 to~\$1.
%Moreover, we observed that the increased transaction activity, over~99\% attributed to the minting of new inscription-bases meme tokens, positively affected other users of zkSync Era by lowering gas fees. Unlike L1 blockchains, ZK rollups experience lower gas fees with increased transaction volumes.

\subsection{Research Questions}
This work is the first empirical study on blockchain performance and user behavior during the inscription boom on EVM chains. We focus primarily on the analysis of the zkSync Era, but we plan to expand to other chains in the final version of our study by adding Arbitrum, Ethereum Mainnet, and other chains. In particular, we focus on the following research questions:

\point{RQ 1: Causes of Transaction Spike}
\textbf{What caused the surge in transactions on EVM-chains at the end of 2023?} We aim to investigate whether the increase in transactions was influenced by the Fear of Missing Out (FOMO) effect, where users speculated on potential profits from transactions, anticipating selling them later. Hence, we examine the transfers, listing, and subsequent selling of transactions post-creation and claim by users. This RQ also leads us to understand how these fees affect user behavior on chain.

During the preliminary studies of inscriptions on zkSync Era, we observed that the spikes in transaction amounts were caused by minting new inscription-based meme coins. Over 80\% of all transactions during the spikes were related to inscriptions, from which 99\% were minting transactions. For a few hours on December 17, over 96\% of all transactions at zkSync Era were related to minting \stress{zrc-20 sync} token.

%We observed some trading activities for two tokens \stress{zrc-20 sync} and \stress{era-20 bgnt}, with \num{400} and \num{1200} transfer (including buy) transactions, and the price oscillating between ~\$0.1 to~\$1 without the trend.

\point{RQ 2: Impact on EVM Blockchain Performance and Cost Efficiency}
\textbf{What impact did the transactions have on the security and scalability of EVM-chains?} Given that certain Layer 2 (L2) solutions experienced downtime, and a significant portion of transactions during late~2023 were inscriptions, we seek to evaluate the performance of these L2 scaling solutions during this stress-test.

We found that zkSync Era users were positively affected by the surge in transaction activities, which is a result of its ZK rollups design. With more L2 transactions compressed into the transaction batch, the median costs of L2 transactions decreased. However, delays in the transaction execution and, consequently, finality were also noted.
 
\point{RQ 3: Speculative Purchases}
\textbf{What is the motivation of users to purchase inscription-bases meme-coins in comparison to ERC-20 meme-coins?} 
%\textbf{Are individuals purchasing in anticipation of transaction value growth?}
Inscriptions and the BRC-20 standard were initially introduced on the Bitcoin blockchain, as this network does not support smart contracts and the creation of ERC-20 tokens. In 2023, inscription-base tokens were minted on the EVM blockchains that support ERC-20 standards. We investigate the motivation of users to create and trade inscription-based tokens on EVM chains.

We observed that the minting process finished for three tokens: \stress{zrc-20 sync}, \stress{era-20 bgnt} and \stress{layer2-20 \$L2}, which are traded on zkSync Era for a fraction of the costs compared to the Ethereum and Bitcoin networks. Even with observed prices in the range of~\$0.1 to~\$1, the costs of minting and listing are covered for the seller. The trading activity for \stress{era-20 bgnt} token continues after the initial inscription boom.

%We also observed that minting on zkSync Era allows for trading at a fraction of the costs compared to the Bitcoin or Ethereum networks. However, trading activity remains limited, with users purchasing one of the two fully minted meme tokens: \stress{bgnt} and \stress{sync} for prices in the range of~\$0.1 to~\$1.

% \point{RQ 4: Evidence of Financial Gain}
% \textbf{Is there evidence of users realizing financial gains from transactions?} We evaluate how the price trajectory of inscription-based meme-coins evolved over time. 

% We observed the limited trading activities for two inscription-bases meme-coins on zkSync Era: \stress{bgnt} and \stress{sync} with prices in the range of~\$0.1 to~\$1 without a trend.

\point{RQ 4: Impact of blobs on inscription security}
\textbf{Does Data Availability of blobs affect inscription ownership?} As inscriptions are appended to transactions as arbitrary data within the input call data, there is no assurance of perpetual availability of this data in the future. We aim to explore how this may impact inscription ownership over time.

We raised security concerns related to the applications of blobs for inscriptions. Blobs, temporary data units introduced by Ethereum's Dencun upgrade~\cite{2024EthereumRoadmap} are designed to enhance rollup scalability and disappear after 18 days. Rollups are not required to store permanent transaction input data, which inscriptions utilize, leaving users to rely solely on the off-chain indexers run by the inscription creators. 

% ====== Contributions ======
\subsection{Summary of Contributions}
%This work is the first empirical study on rollups in general and the first one on inscriptions on EVM-compatible chains. We investigated inscriptions' impact on the blockchains' security and scalability during the inscription boom at late 2023. 
We summarize our contributions as the following: 

\paraib{Impact on rollup users}
We observed that zkSync Era users were positively affected by the surge in transaction activities, which is a result of its ZK rollup design. With more L2 transactions compressed into the transaction batch, the median costs of L2 transactions decreased. %However, delays in the transaction execution and, consequently, finality were also noted.

\paraib{Security assumptions for rollups.}
We raised security concerns related to the applications of blobs for inscriptions. %Blobs are temporary data units introduced on March~13\tsup{th}, 2024, by Ethereum's Dencun upgrade~\cite{2024EthereumRoadmap}, designed to enhance rollup scalability. 
While blobs disappear after 18 days, rollups are not required to store permanent transaction input data, which inscriptions utilize, leaving users to rely solely on the off-chain indexers. 

%\paraib{Inscription Minting and Trading}
%We found that the increased transaction activity was in~99\% attributed to the minting of new inscription-bases memecoins.
%We also observed that minting on zkSync Era allows for trading at a fraction of the costs compared to the Bitcoin or Ethereum networks. However, trading activity remains limited, with users purchasing one of the two fully minted meme tokens: \stress{bgnt} and \stress{sync} for prices in the range of~\$0.1 to~\$1.

\paraib{Inscription data characterization.}
We employ a data-driven approach to analyze zkSync Era, which experienced significant transaction activities in~2023. Our analysis finds these transactions were related to minting inscription-based new ERC-20-type tokens (~99\%). These tokens were later traded at a fraction of the costs compared to inscriptions on the Bitcoin or Ethereum networks, with prices in the range of~\$0.1 to~\$1.
%Our anlysis includes measuring the frequency of inscription minting, the number of successful addresses claiming inscriptions, total supply, fees spent per address during minting, and the distribution of inscriptions per address. 
We found that inscription standards on zkSync Era are based on BRC-20 standards with some modifications to facilitate trading and listing at NFT marketplaces. %and mostly applied to minting new ERC-20-type tokens. % \REVISIT{As the minting process of the declared total circulating supply of these tokens has not yet finished, trading and transfer activities remain limited.}
% John: For BigInt and ZKS market, the top two inscriptions market place on zkSync, they have already finished and reached the total supply claming amount in a matter of max. 1--2 days. 

\paraib{Scientific reproducibility.}
To promote scientific reproducibility, we commit to making both our datasets and scripts publicly available.

%% file: sections/related_work.tex
\section{Related work}
\label{sec:related-work}

%In this section, we present the relevant work in the literature regarding ordinals and inscriptions.

\begin{comment}
We review the literature related to ordinals and inscriptions below.
\begin{itemize}
    \item Bridging BRC-20 to Ethereum~\cite{yu2023bridging}
    \item Understanding BRC-20: Hope or hype~\cite{wang2023understanding}
    \item BRC-20: Hope or Hype~\cite{wang2023brc}
    \item Bitcoin Ordinals: Determinants and Impact on Total Transaction Fees~\cite{bertucci2023bitcoin}
    \item BRC-20 Tokens: A Primer~\cite{BRC-20@Binance}
    \item The first ever cross-chain \href{https://twitter.com/inscribe_app/status/1744444280129655098}{marketplace for inscriptions}, automated WebApp \& Bot and Multichain Indexer API for every chain (Bitcoin ecosystem  + EVM/Non-EVM).
\end{itemize}
\end{comment}

Below we review several papers that are concerned with \emph{inscriptions} and \emph{ordinals}. 
Wang et al.~\cite{wang2023brc,wang2023understanding} conducted a comprehensive analysis of BRC-20 tokens within the Bitcoin network. Their investigation revealed a notable surge in interest and user participation during the BRC-20 movement.
The study involved a comparison between these BRCs and the widely-used Ethereum ERC-20 tokens, employing metrics such as average price return, volatility, and other relevant factors. The findings of the research yielded intriguing insights. %\REVIEW{KG: We should calculate the same metrics: average price return, volatility, etc. The prices need to come from the NFT/Inscription marketplaces}
For example, the study highlighted a significant influx of users into the BRC market within a month, coupled with a subsequent decline in enthusiasm over the following months. In essence, users displayed a tendency to lose interest after the initial surge. 
In the broader context, the research emphasized that, despite the hype, BRC-based tokens constituted only a modest fraction of the overall market size when juxtaposed with ERC-like tokens on the Ethereum platform.
Additionally,~\cite{yu2023bridging} proposed a bridging mechanism of BRC-20 tokens from the Bitcoin network to the Ethereum blockchain, and~\cite{bertucci2023bitcoin} analyze the impact of ordinals on transaction fees in Bitcoin.

While these previous studies have explored ordinals in Bitcoin, we are not aware of research concerning inscriptions. Therefore, our investigation aims to fill this gap by analyzing inscriptions activities, particularly those found within Layer 2s, in order to comprehend their processes of minting, transfer, purchase, and sale. To our knowledge, this study is the first to conduct an in-depth analysis of inscriptions and the user dynamics that arose from the inscription boom.

%% file: sections/background.tex
\section{Background}\label{sec:background}
Addressing the scalability challenges and gas price fluctuations in blockchain networks involves two primary approaches: Layer-1 (L1) scaling and Layer-2 (L2) scaling. This section provides a short introduction to rollups - non-custodial Layer-2 (L2) blockchains.

\parai{Layer-1 Scaling (L1).}
This approach involves creating an entirely new blockchain with unique consensus mechanisms and dedicated physical infrastructure to maintain the network.

\parai{Layer-2 Scaling (L2).}
In contrast, L2 scaling adopts a different strategy. It involves executing computations outside the main blockchain, with the results or final state subsequently recorded on the underlying chain. Major types of L2 solutions include state (payment) channels, plasma, and rollups. While plasma and state channels focus on moving both data and computation off-chain, rollups move computation and state storage off-chain, retaining compressed data per transaction on the underlying chain.

\parai{Rollups.}
They function as a form of L2 scaling by conducting computations off-chain and storing the results on the main chain. Essentially, they operate like blockchains, generating blocks and then recording snapshots of these blocks on the primary chain. However, in the rollup environment, operators are not inherently trusted. This lack of trust implies that operators have the potential to behave adversarial by halting block production, generating invalid blocks, withholding data, or engaging in other malicious activities. Therefore, they typically implement a security routine to guarantee security.

\begin{figure}[!ht]
  \centering
        \includegraphics[width=0.9\linewidth]{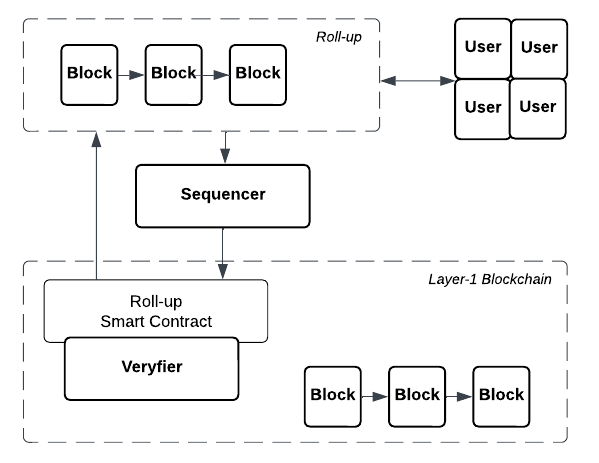} 
    \caption{General Architecture of rollups -- Non-custodial L2 blockchain scaling solution}
    \label{fig:rollup}
\end{figure}

Ensuring the correctness of the state in rollups involves two main approaches: optimistic and zero-knowledge proofs (ZKP). Some rollups, such as zkSync Era, employ Zero Knowledge (ZK) proofs. These proofs involve computing cryptographic proofs that validate the correctness of the computation. On the other hand, others, like Arbitrum, adopt an optimistic approach that assumes all statements are true unless contested by someone in the network.
Figure~\ref{fig:rollup} provides an illustration of the rollup architecture, highlighting key components such as sequencers and verifiers.

\parai{Sequencers.}
These entities are responsible for rolling up transactions to the L1 chain. By bundling transactions, rollups effectively manage to achieve cost savings in terms of gas fees.

\parai{Verifiers}
Operating as smart contracts on the L1 chain, veryfiers play a crucial role in validating the transactions stored by the sequencer. They ensure the correctness of the transactions, thereby maintaining the integrity of the overall system~\cite{sguanci2021layer, yee2022shades, gangwal2022survey}.

%--------------------------------------------------------------------------------------------------------------
\section{Inscriptions and Ordinals}

Inscriptions and ordinals involve the recording of arbitrary data on the blockchain. On EVM-chains, users encode hex data into the transaction input call data (refer to columns \stress{JSON} in Table~\ref{tab:inscriptions}), usually also setting the transaction's \stress{from} and \stress{to} attributes with the same user's addresses. This structure resembles a self-transfer made from a user to their own address. 

\subsection{Operation Types}
There are various protocol standards that define the structure of the inscription. \stress{BRC-20} was the first standard proposed for Bitcoin, and it includes three main types of operations: \stress{deploy}, \stress{mint}, and \stress{transfer}, each encrypted into a single transaction~\cite{wang2023brc}. The protocol standards on EVM chains extend the set of operations, e.g., \stress{layer2-20}: claim, \stress{era-20}: \stress{list} and \stress{buy}, \stress{era-20}:\stress{sell}.

\begin{table*}[tb]
\centering
\begin{tabular}{cccrrl}
\toprule
\thead{Protocol} & \thead{Tick} & \thead{Operation} & \thead{Total Supply} & \thead{Limit}    & \thead{JSON data}    \\
\midrule
    zrc-20 & sync  & deploy & $21\times10^6$ & \num{4} & \scriptsize{\verb|{"p":"zrc-20","op":"deploy","tick":"sync","amt":"21000000","limit":"4"}|}\\
    zrc-20 & sync  & transfer & --- & \num{4} & \tiny{\verb|{'p': 'zrc-20', 'op': 'transfer', 'tick': 'sync', 'amt': '4'}|}\\
    zrc-20 & sync  & mint & --- & \num{4} & \tiny{\verb|data:,{"p":"zrc-20","op":"mint","tick":"sync","amt":"4"}|}\\
    zrc-20 & zksi  & deploy & $21\times10^8$ & $1\times10^4$ & \tiny{\verb|{"p":"zrc-20","op":"deploy","tick":"zksi","max":"2100000000","limt":"10000"}|}\\
    zrc-20 & zkss  & deploy & $21\times10^6$ & $1\times10^3$ & \tiny{\verb|{"p":"zrc-20","op":"deploy","tick":"zkss","max":"21000000","lim":"1000"}|}\\
    zrc-20 & zkss  & sell & --- & \num{1000} & \tiny{\verb|data:,{"p":"zrc-20","op":"sell","orderMessage":[{"amt":10000,"nonce":"62...",|}\\
    --- & ---  & --- & --- & --- & \tiny{\verb|"seller":"0x9b...","sign":"0xc5d3...","tick":"zkss","value":0.015}]}|}\\
    zrc-20 & zkzk  & deploy & $21\times10^6$ & $21\times10^6$ & \tiny{\verb|{"p":"zks-20","op":"deploy","tick":"zkzk","max":"21000000","lim":"21000000"}|}\\
    era-20 & bgnt  & deploy & $21\times10^6$ & \num{5} & \scriptsize{\verb|{"p":"era-20","op":"deploy","tick":"bgnt","max":"21000000","lim": "5"}|}\\
    era-20 & bgnt  & deploy & $21\times10^6$ & \num{5} & \scriptsize{\verb|{"p":"era-20","op":"deploy","tick":"bgnt","max":"21000000","lim": "5"}|}\\
    era-20 & bgnt  & list & --- & \num{5} & \tiny{\verb|data:,{"p":"era-20","op":"list","tick":"bgnt","amt":"250","price":"1500000000000000"}|}\\
    era-20 & bgnt  & buy & --- & --- & \tiny{\verb|data:,{"p":"era-20","op":"buy","tx":"0xda..."}|}\\
    layer-2 & \$L2  & claim & --- & \num{1000} & \tiny{\verb|data:,{"p":"layer2-20","op":"claim","tick":"$L2","amt":"1000"}|}\\
%    XXXX & XXXX  & XXXX & XXXX & XXXX & XXXX\\
\bottomrule
\end{tabular}
\caption{Example of inscriptions data recorded on-chain: protocol name, operation, token name (Tick), total supply and maximum amount of tokens minted each round (Limit)}
\label{tab:inscriptions}
\end{table*}

\parai{Deploy.}
The \stress{deploy} action specifies the protocol name, token tick, total supply, and the maximum amount of tokens a user can mint (or claim) per transaction. Table~\ref{tab:inscriptions} displays the on-chain inscriptions data. For instance, to deploy an inscription, a transaction containing a deploy action should be recorded on-chain, marking the initiation of an inscription event. In the case of the zrc-20 sync inscription, the protocol specifies a total supply of 21 million inscriptions and a limit of 4 tokens per transaction.

\parai{Mint.}
After the transaction deploying the inscription is persisted to the chain, users can issue a \stress{mint} action to actually mint (or in this case claim ownership) of the tokens. To initiate this, users need to issue a transaction with an input call data encoded to HEX code, specifying the protocol and tick that jointly identify the inscription-based token. In the same transaction, users provide the number of tokens they want to claim (refer to column \stress{Limit} in Table~\ref{tab:inscriptions}) that must be not higher than the maximum limit specified in \stress{deploy} operation for the given protocol-tick pair. The off-chain \stress{indexer} is responsible for assuring the integrity of inscription-based tokens.

% For example, the first deploy op we received on zkSync that corresponds to {"p":"zrc-20","op":"deploy","tick":"sync","amt":"21000000","limit":"4"} was at block number 21636531 and transaction 0xd50b3 ... ec7f0. From this point, the mint event started.

\parai{Transfer.} Allows to transfer the ownership of the inscription token from one address to the other. The inscription token is identified by protocol standard and token name (tick). Tokens with the same names (ticks) can be deployed multiple times using various protocol standards.

\parai{List.} Allows to list the inscription on the NFT marketplace; it specifies the amount of tokens and their price in Ether.

\parai{Buy.} Used to buy inscription tokens from NFT marketplaces, it points to the transaction that listed the inscription on the marketplace. The \stress{buy} operation specifies only protocol standards, as token name (tick), amount, and price are declared in the \stress{list} operation.

\parai{Sell.} Used to sell inscription, it specifies the protocols and the order message with seller and signer details as well as the amount of sold tokens and their prices.

\parai{Claim.} Similarly to mint, it allows to mint and claim ownership of a new token. This operation is used within \stress{layer2-20} standard to mint a new inscription token \stress{\$L2}. The \stress{\$L2} token is a multi-layer2 inscription token that its creator allows to mint on multiple blockchains. However, there is just one  \stress{deploy} operation defining the token max supply. The off-chain \stress{indexer} of the creator is responsible for assuring the integrity of data in these circumstances.
% What do we mean by other L2 BCs? KG: Other then zkSync Era. L2 token is the 2nd largest in terms of transactions and it does not have a deploy on zkSync Era. It is meme-coin across various rollups. 

\subsection{Comparison with NFT and ERC-20}
\parai{Comparison with NFTs}
Within the Ethereum blockchain, non-fungible tokens (NFTs)~\cite{wang2021non} are instantiated via smart contracts. Each user is allocated a distinctive token ID, enabling them to assert ownership over a particular asset. These assets, including JPEG files or CryptoPunk images, are stored off-chain on a server.

Unlike NFTs, inscription-based tokens do not depend on any smart contract and, thus, do not allow upgrades. The link to the asset file is inscribed into the transaction data. % due to the amount of data required to mint the inscription, the daily total minting limits exist.  
In a blockchain with a maximal native token supply, such as Bitcoin, the amount of inscription is bound by the blockchain network limits. In contrast, NFTs based on smart contracts lack such limitations, theoretically allowing for unlimited minting. Also, each inscription is allocated a position in the blockchain, which creates the opportunity to derive the additional value of the inscription from the location within the block.

\parai{Comparison with ERC-20}
The other application of inscriptions is to create ERC-20-style tokens. As inscription-based tokens are not reliant on smart contracts, they are not susceptible to the risk of smart contract upgrades, and their maximal circulating supply is declared once in the \stress{deploy} operation. The minting and transfer of each new inscription-bases token can be tracked directly on the blockchain and cannot be altered.

Nevertheless, due to the lack of support for smart contracts, inscription-based tokens offer presently minimal utility. They are predominantly used for speculation and belong to the category of meme-coins~\cite{wang2023brc}. Based on the inscription technology, it is possible to mint a single NFT-style token or a group of tokens with a predefined token supply in circulation. However, it is the off-chain indexer, operated by the inscriptions' creator, that is responsible for assuring the data integrity of minted tokens. 
Trading of inscription-based tokens occurs at NFT marketplaces, as they are not compatible with ERC-20 standard required by decentralized exchanges~\cite{xu2023sok}. 

%% file: sections/dataset.tex
\section{Data collection}
\label{sec:dataset}

\begin{table*}[tb]
\centering
\scriptsize
\begin{tabular}{cccccccc}
\toprule
\setlength{\tabcolsep}{6pt}
\thead{Chain} & \thead{Start date} & \thead{End date} & \thead{\# of issuers} & \thead{\# of blocks}    & \thead{\# of transactions}  & \thead{block range}  &
\thead{\# of inscriptions}  \\
\midrule
    zkSync Era & June 18\tsup{th}, 2023 & Mar. 25\tsup{th}, 2024  & \num{6768715} & \num{23070883} & \num{289466141} & \num{6332862}--\num{29799866} &
    \num{17054466}\\
\bottomrule
\end{tabular}
\caption{Dataset used to analyze inscription events.}
\label{tab:dataset}
\end{table*}

We employ a data-driven approach in our study, specifically utilizing blockchain data obtained from an archive node and a Web3-compatible API for zkSync Era. These data sources enable us to collect comprehensive data containing all information recorded on evaluated chains. This includes data about blocks, transactions, and events (or logs triggered during the execution of smart contracts) specifically related to inscriptions. Table~\ref{tab:dataset} provides an overview of the specifics of our dataset.

It is important to note that our focus is solely on blockchain data associated with inscriptions. While a significant portion of inscriptions involves the issuance of self-transfer transactions (i.e., where issuers initiate a transfer to their own address), we adopt a broader perspective. This involves considering instances where inscriptions are added to the chain, irrespective of whether through a self-transfer or not. This is important to capture the total fees users spent with inscriptions.

To identify these inscriptions on-chain, we specifically search for transactions with input call data starting with ``\verb|0x646174613a|'', representing ``\verb|data:|'' in ASCII. This criteria establishes crucial conditions for recognizing inscriptions within the blockchain.

In total, we found \num{17054466} transactions containing inscriptions in our data set issued by~\num{470864} addresses and added to~\num{2713534} blocks.
The average number of inscriptions per block is~\num{6.28} inscriptions. We rely on this data set to conduct our empirical analysis of inscriptions.

%% file: sections/inscriptions-characterization.tex
%\section{Inscriptions characterization}\label{sec:characterization}
%In this section we leverage the data set to characterize the inscriptions recorded on zkSync Era, Arbitrum, Optimism, and Ethereum.
%\color{blue}
%Title: Roll-up evaluations
%Methodology: We analyze data from X to Y on roll-ups X, Y, Z. We choose rollups X, Y, Z, because...
%Why not Polygon? Polygon got spike in gas fees
%To calcualted: Time and fees.
%Impact on other users.
%\color{black}

\section{Empricial Study of Inscriptions on zkSync Era}
\label{subsec:zksync}

We started the primary analysis with zkSync Era, EVM-compatible ZK rollup. 
%It allows developers to make code developed to Ethereum mainnet portable or compatible with zkSync Era with minimum changes required. 
This rollup is among the most used rollups with an estimated total value locked (TVL) of about \$171.1mn in DeFi~\cite{2024DeFiLlama}. Its daily transactions' activity ranges from~\$1 million to~\$1.5 million~\cite{2024DuneZKSync}. For this reason, zkSync Era is among the rollups we chose as a subject of our study.

\begin{figure}[!ht]
  \centering
        \includegraphics[width=0.9\linewidth]{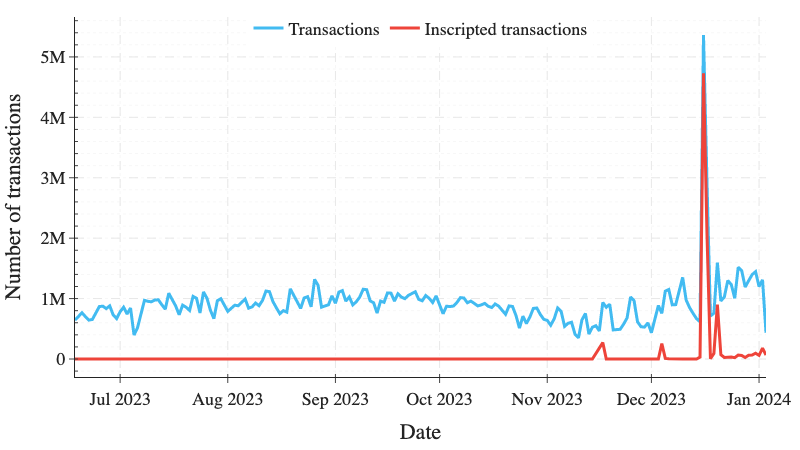} 
    \caption{Number of inscriptions in comparison to the total number of transactions per day on zkSync Era
}
    \label{fig:transactions}
\end{figure}

\begin{figure}[!ht]
  \centering
        \includegraphics[width=0.9\linewidth]{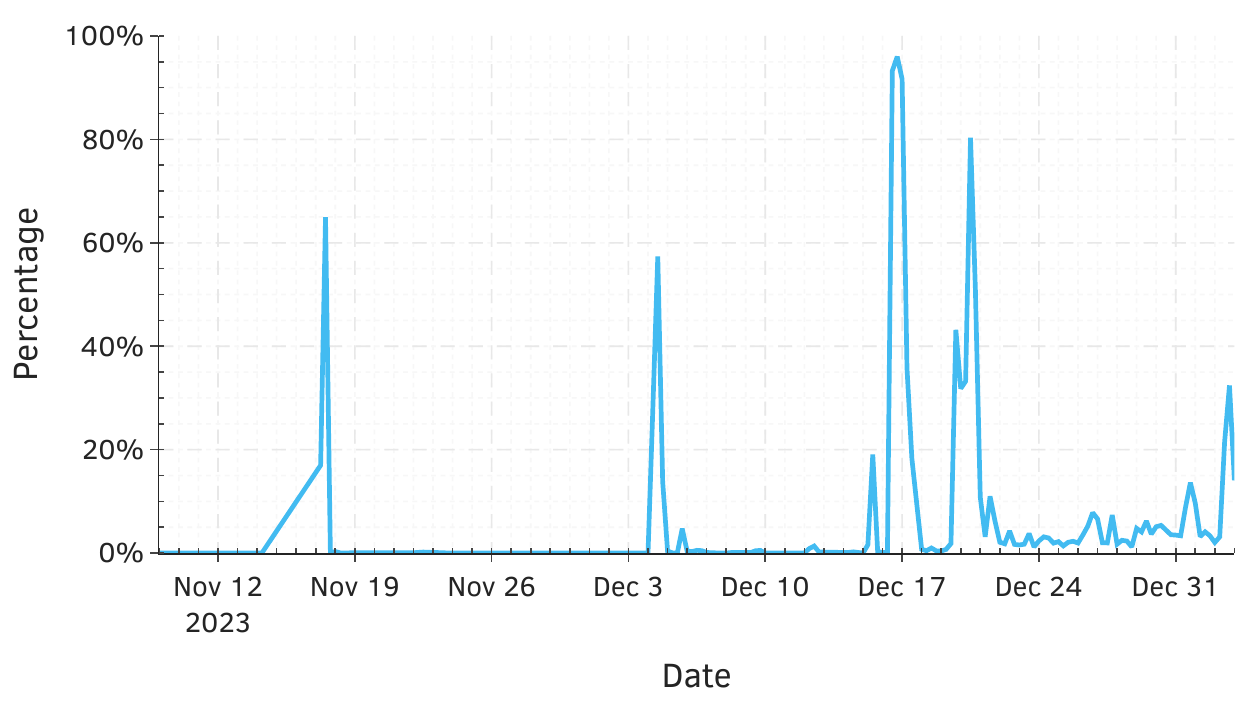} 
    \caption{Percentage transactions with inscriptions in comparison to the total number of transactions per day on zkSync Era}
    \label{fig:fraction}
\end{figure}

\parai{Overall transactions.} During November and December~2023, inscriptions led to spikes in the overall number of transactions in zkSync Era, as depicted in figures~\ref{fig:transactions} and~\ref{fig:fraction}. Notably, on December~17, the overall number of daily transactions grew to almost~\$5 million, with inscription-related transactions representing over~95\% of daily transactions. Spikes in inscription transactions could also be observed on November~17, December~5, December~17, December~20, when overall inscriptions represented over~60\% of all transactions. 

%1) Transaction analysis
%a) First, there was that amount of transaction.
%b) Second, this percentage of this transaction was inscriptions.
%c) Impact on other users (fees analysis) - show gas fees costs

\begin{figure}[!ht]
  \centering
        \includegraphics[width=0.9\linewidth]{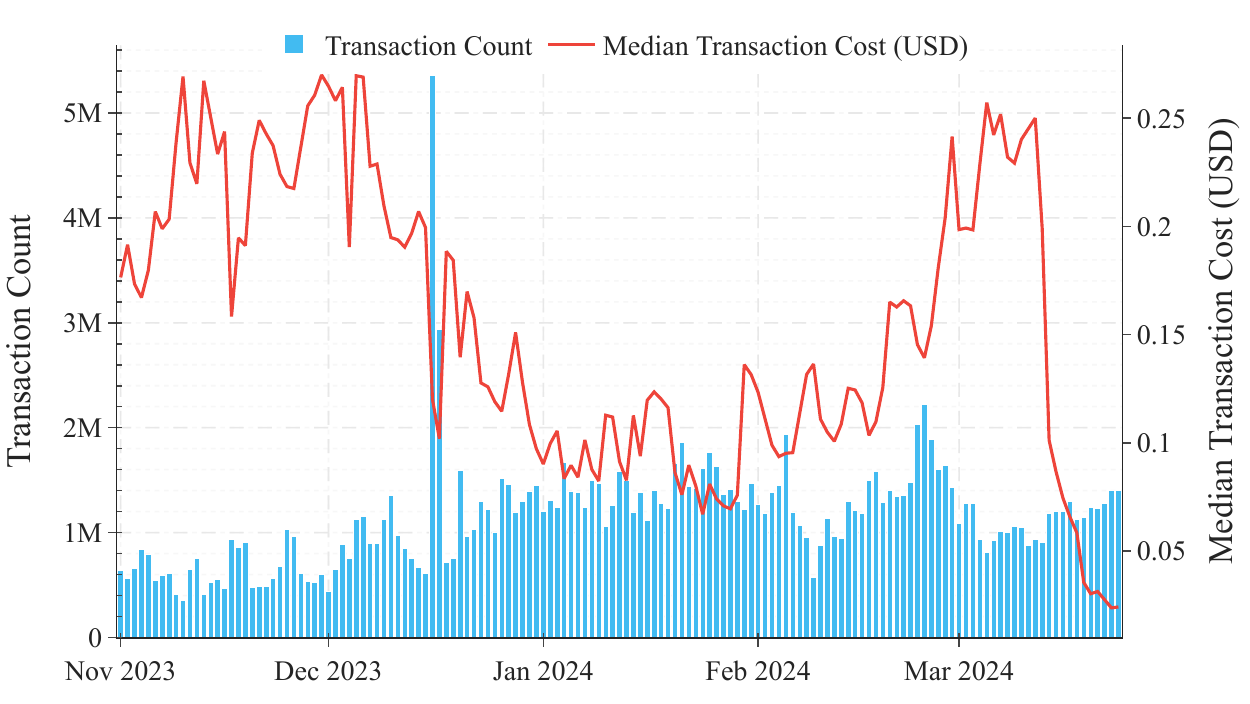} 
    \caption{Daily transaction count and median transaction cost paid by the user of zkSync Era.}
    \label{fig:gasfees}
\end{figure}

\parai{Gas Fees}
It can be further observed that the spike in daily transaction counts on the mentioned days led to a decrease in median gas fees per transaction for other users, as depicted in Figure~\ref{fig:gasfees}. An inherent characteristic specific to ZK rollups, such as zkSync Era, is their capability to distribute the cost of generating validity proofs across many transactions. ZK rollups rely on ZK-STARKs (Zero-Knowledge Scalable Transparent Argument of Knowledge)~---~cryptographic methods~---~to validate transactions off-chain and submit them in batches to underlying L1 blockchains. With more transactions in a batch, the cost per transaction decreases. Consequently, unlike at L1 blockchains, gas cost at zk-rollups is reduced with a larger transaction volume, which occurred on November~17, December~5, December~17, December~20 on zkSync Era. The drop of transaction costs on~13th March is caused by Ethereum's Dencun upgrade~\cite{2024EthereumRoadmap} that introduced blobs and danksharding designed to increase the efficiency of Ethereum rollups.

\begin{figure}[!ht]
  \centering
        \includegraphics[width=0.9\linewidth]{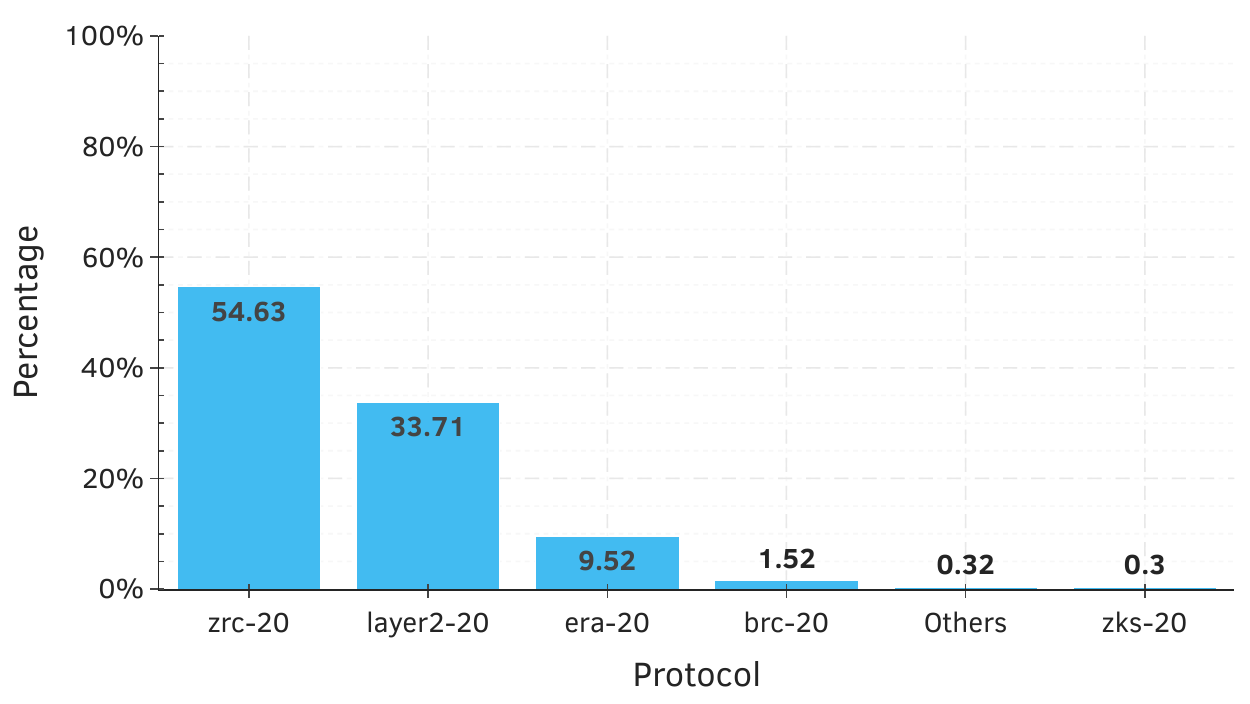} 
    \caption{Top 5 inscription protocols on zkSync Era}
    \label{fig:top15protocol}
\end{figure}

\begin{figure}[!ht]
  \centering
        \includegraphics[width=0.9\linewidth]{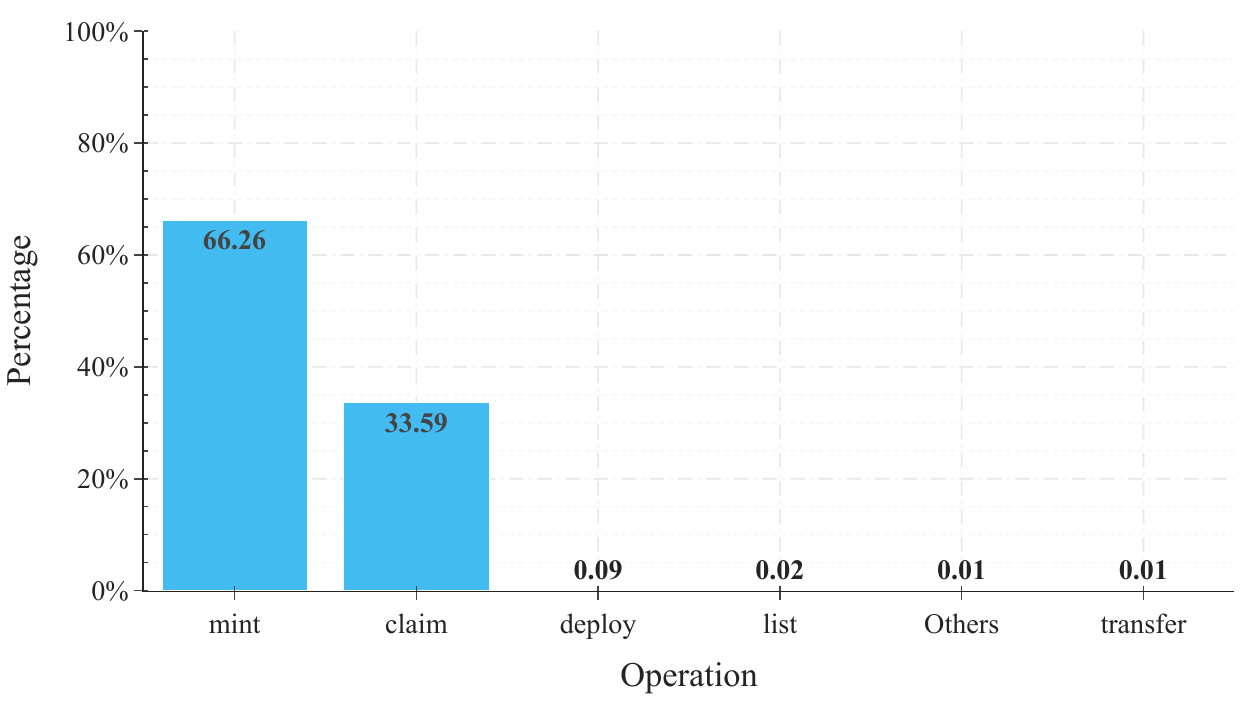} 
    \caption{Top 5 inscription operations on zkSync Era}
    \label{fig:top15operator}
\end{figure}

\parai{Protocols and Operations}
During the analyzed period from June~2023 to March~2024, there were \num{17054466} transactions deploying inscriptions on zkSync Era: \num{196763} for minting NFT-type ordinals and \num{16863729} related to meme coins and transfers. Fig \ref{fig:top15protocol} depicts the breakdown of inscription protocol standards, and fig \ref{fig:top15operator} the operations within these standards. The dominant inscription standards were \stress{zrc-20 (over half of the transaction)}, \stress{layer2-20} and \stress{era-20}, as for the operation~---~\stress{mint} and \stress{claim} represented over~99\% of inscription-related transactions. It was also observed that these standards contain new operations types, such as \stress{claim}, \stress{list}, \stress{buy} and \stress{sell} that are not present in the original \stress{brc-20} implementation on Bitcoin.

%As per Table~\ref{tab:inscriptions}, these are inscriptions that set the operation command (\stress{op}) to \stress{deploy}, meaning a new inscription is available on-chain \MISSING{number of deploy}. We also identified \num{8967202} transactions used to mint the new available inscriptions. Some inscriptions, \num{186720} in total, used a different operator to mint the tokens called \stress{claim}. There are other operators used, as for example, transfer (\num{609}), list (\num{141}), sell (\num{103}), and buy (\num{42}).

\begin{figure}[!ht]
  \centering
        \includegraphics[width=0.9\linewidth]{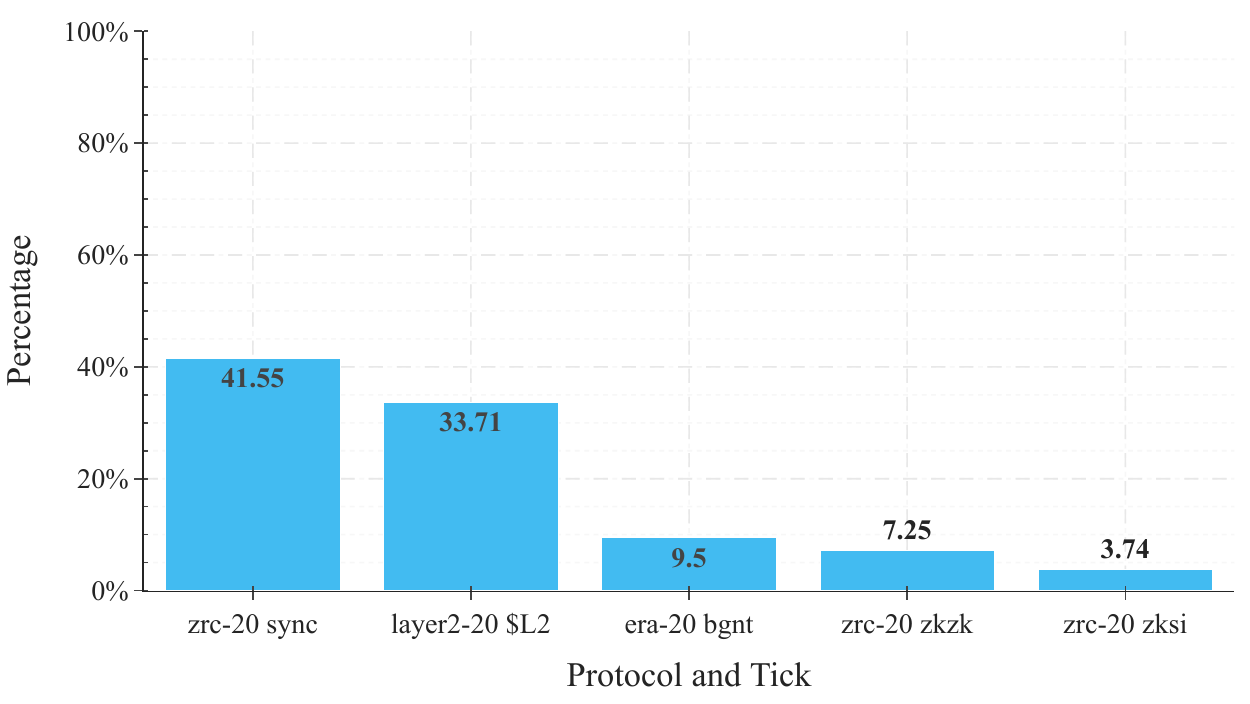} 
    \caption{Top 5 inscription tokens on zkSync Era}
    \label{fig:top15tick}
\end{figure}

\parai{Tokens and operations}
%Each inscription-based token is uniquely defined by the protocol standard and the tick in the first \stress{deploy} operation registered on the blockchain. 
The Fig~\ref{fig:top15tick} presents tokens (ticks) per standard, with \stress{zrc-20 sync}, \stress{layer2-20 \$L2 }, \stress{erc-20 bgnt} and \stress{zrc-20 zksi} representing over~90\% inscription transactions. Interestingly, we observed that other protocols also deploy and mint tokens with previously used ticks, e.g., \num{17757} transactions with \stress{sync} tick at non-\stress{zrc-20} protocols, \num{4909} transactions with \stress{bgnt} tick at non-\stress{era-20} protocols. The \stress{deploy} operation for \stress{layer2-20 \$L2} token was not registered on the zkSync Era, \stress{layer2-20 \$L2} is a token minted and traded across multiple L2 with off-chain indexer assuring its data integrity.

\begin{figure}[!ht]
  \centering
        \includegraphics[width=\linewidth]{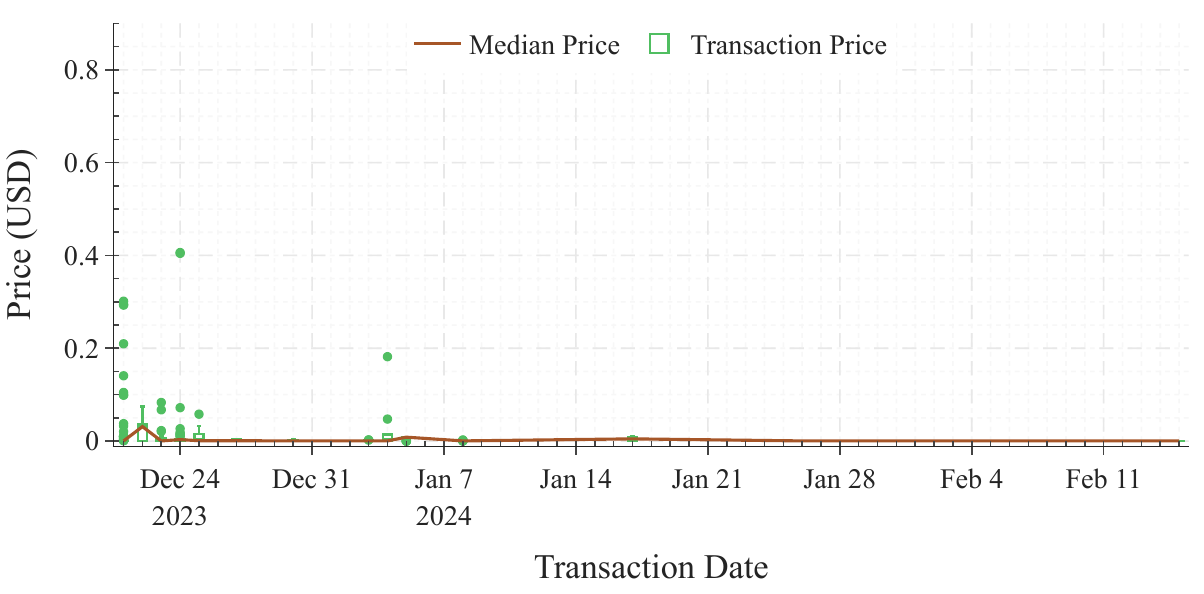} 
    \caption{Price per token at every transaction of \stress{zrc-20 sync} token}
    \label{fig:transfersync}
\end{figure}

\begin{figure*}[!ht]
  \centering
        \includegraphics[width=\linewidth]{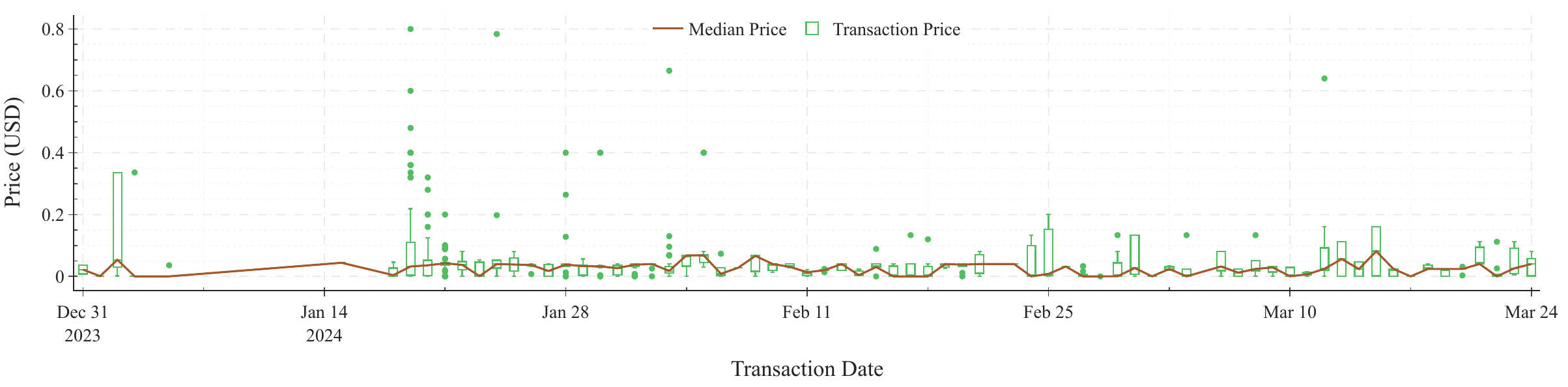} 
    \caption{Price per token at every buy transaction of the \stress{era-20 bgnt} token}
    \label{fig:buybgnt}
\end{figure*}

\parai{Inscription Trading} 
We observed \num{3246} of \stress{list} transactions, followed by \num{1314} - \stress{transfer}, \num{1206} - \stress{buy} and \num{103} - \stress{sell} operations. Most of these transactions are related to two memecoins: \stress{zrc-20 sync} and \stress{era-20 bgnt}. We found \num{477} of \stress{transfer} transactions for \stress{zrc-20 sync}, with prices per token depicted in Figure~\ref{fig:transfersync}, and \num{1148} of \stress{buy} transactions of \stress{era-20 bgnt}, with prices in Figure~\ref{fig:buybgnt}.

During the studied period, the price of \stress{zrc-20 sync} and \stress{era-20 bgnt} oscillated in the ranges of~\$0.1 to~\$4 and~\$0.1 to~\$8 respectively. As these trades occurred in batches of 4 and 5 tokens, they provided minimum compensation to the seller to cover the gas fees necessary to mint the inscription. Trading of inscriptions at such low prices with gas fees compensated is possible thanks to the low gas fees on zkSync Era, in the range of~\$0.05~--~\$0.25 during the period \ref{fig:gasfees}. For \stress{zrc-20 sync}, we observed minimum values were close to 0 and a maximum of \num{0.406307}\$ with average and median \num{0.00781}\$ and \num{0.000403}\$ respectively and std \num{0.043483}. For \stress{era-20 bgnt}, trading occurred in the range of 0 and \num{0.8}\$, average and median values: \num{0.031188}\$ and \num{0.042997}\$ and std \num{0.078061}. As depicted in Fig \ref{fig:buybgnt}, trading of \stress{era-20 bgnt} token continues from its minting throughout the whole studied period.

%The most traded tokens are depicted in the appendix in fig \ref{fig:top5transfer}, with over half of the inscription trades related to \stress{zrc-20 sync}. The low trading activities of inscription-based tokens can be attributed to the unfinished minting process. None of the tokens reached the maximum circulation supply (\num{21000000} in most cases) and can still be minted for the cost of the gas fees on the rollups, as depicted in the appendix in fig \ref{fig:top5mint}. Furthermore, most of the observed transfers were transfers of ownership between wallets without corresponding payment tracked on-chain.

%c) What was happening with inscriptions - these were deployment and minting transactions, an overview of these minting events (total costs, how long it was, amount of tokens)

%3) Symbil and transaction relations between inscptions, transfer patters.

%4) Conclusions Avoid judgments. This was a test of projects that involved a large amount of transactions, And from the roll-ups analyzed, it is shown that the ZkSync Era proved the most salable. Explain high level why - changes in prover

%% file: sections/discussion.tex
\section{Discussion}
\label{sec:discussion}

\parai{Inscriptions Security}
On March 13, 2024, Ethereum underwent the Dencun upgrade~\cite{2024EthereumRoadmap}, which introduced temporary data storage - \stress{blobs}. Blobs, initially designed to increase the efficiency of Ethereum rollups, were used to issue new inscription directly in Ethereum blockchains, called \stress{BlobScriptions}~\cite{2024DuneEVMInsctiptions}. Yet, blob data is intended to be stored by Ethereum nodes for~18 days, during which rollups verify their transactions. After the 18-day period, BlobScriptions disappear from the Ethereum blockchain and are only stored (and available) in the indexer of the creator protocol, outside of the blockchain. Similarly, rollups are not required to store (outside of blobs) the transaction input data used by rollup-based inscriptions, raising questions about the security of inscriptions.

\parai{Inscriptions: Good or Bad?}
A report from Binance on BRC-20 highlights that the unexpected surge in transactions resulted in an increase in transaction fees for many blockchains and rollup~\cite{BRC-20@Binance}. The report suggests that such a rise in fees can be seen as a natural progression in blockchain adoption. To illustrate, if a few million more individuals had chosen to use Bitcoin or other blockchains for peer-to-peer transactions, a similar spike in transaction fees would have occurred. Therefore, attributing this spike solely to BRC-20s is considered irrelevant. More worrisome, other protocols, such as Arbitrum, experienced downtime of approximately~78 minutes~\cite{Tom-Arbitrum@Cointelegraph}. Consequently, this led to users worrying about long transaction inclusion times. This led the blockchain community to debate whether ordinals and inscriptions benefit the blockchain users. 

However, others might argue that existing blockchain protocols were not fully equipped to handle such a significant influx of transactions. As described in this work, zkSync Era encountered higher transaction-per-second (TPS) rate and reduction in gas fees when for a couple of hours, approximately~96\% of all transactions were related to inscriptions. 

\parai{Inscriptions in DeFi}
Inscriptions are not compatible with ERC-20 tokens and consequently can not be traded at decentralized exchanges based on automated market makers such as Uniswap v2, which is the base for trading tokens with low market capitalization. Today, inscriptions are listed at NFT-marketplaces in a limited-order book type, reducing their liquidity. Also, for many inscription-based meme-coins, their minting processes have not yet finished, and new tokens are expected to be minted until the token limit declared in the \stress{deploy} operation is fully researched. The wrapper of inscription-based meme-coins into the ERC-20 token could allow to list of inscriptions at decentralized exchanges and attract traders of meme-coins that are not used to NFT marketplaces.

%Minting a new inscription token is cheaper than purchasing it at the NFT marketplace, as it requires only the gas fee. Consequently, the increase in trading activities (\stress{transfer} operations) can be expected when the maximum circulating supply of the inscription tokens is reached.

%% file: sections/conclusion.tex
\section{Conclusion}
\label{sec:conclusion}

Inscriptions boom was the first stress test for roll-ups, a novel approach for blockchain scalability. It revealed stark differences between various rollup implementations. We observed that the increased transaction activity positively affected users of the zkSync Era, resulting in lowered gas fees. Unlike L1 blockchains, ZK rollups, such as zkSync Era, experience lower gas fees with increased transaction volumes.

We found that the increase in transaction activity was attributed to the minting of new inscription-based meme-coins (over~99\% of inscription transactions)~---~the process that has yet finished for three tokens \stress{zrc-20 sync}, \stress{era-20 bgnt}, \stress{layer2-20 \$L2} and results in limited trading activity. Compared to the Bitcoin and Ethereum blockchains, zkSync Era network allowed for inscription trading at factional prices, which is not possible on L1 networks due to high gas fees.

Inscriptions, initially proposed as a walk-around for the lack of virtual machines on the Bitcoin blockchain, proved their applicability in EVM blockchains. Nevertheless, the introduction of temporary data storage on Ethereum~---~blobs~---~and rollups' dependence on them~---~puts the security and longevity of blob- and rollup-based inscription into question.

%% file: sections/acknowledgments.tex
%\section*{Acknowledgments}

%% file: main.bbl

\begin{thebibliography}{31}

%%% the Web address.

\ifx \showCODEN    \undefined \def \showCODEN     #1{\unskip}     \fi
\ifx \showDOI      \undefined \def \showDOI       #1{#1}\fi
\ifx \showISBNx    \undefined \def \showISBNx     #1{\unskip}     \fi
\ifx \showISBNxiii \undefined \def \showISBNxiii  #1{\unskip}     \fi
\ifx \showISSN     \undefined \def \showISSN      #1{\unskip}     \fi
\ifx \showLCCN     \undefined \def \showLCCN      #1{\unskip}     \fi
\ifx \shownote     \undefined \def \shownote      #1{#1}          \fi
\ifx \showarticletitle \undefined \def \showarticletitle #1{#1}   \fi
\ifx \showURL      \undefined \def \showURL       {\relax}        \fi
% invisible to TeX
\providecommand\bibfield[2]{#2}
\providecommand\bibinfo[2]{#2}
\providecommand\natexlab[1]{#1}
\providecommand\showeprint[2][]{arXiv:#2}

\bibitem[202(2024a)]%
        {2024DeFiLlama}
 \bibinfo{year}{2024}\natexlab{a}.
\newblock \bibinfo{title}{{DeFi Llama}}.
\newblock
\newblock
\urldef\tempurl%
\url{https://defillama.com/}
\showURL{%
\tempurl}


\bibitem[202(2024b)]%
        {2024DuneEVMInsctiptions}
 \bibinfo{year}{2024}\natexlab{b}.
\newblock \bibinfo{title}{{Dune Dashboard - EVM Inscriptions}}.
\newblock
\newblock
\urldef\tempurl%
\url{https://dune.com/hildobby/inscriptions}
\showURL{%
\tempurl}


\bibitem[202(2024c)]%
        {2024DuneZKSync}
 \bibinfo{year}{2024}\natexlab{c}.
\newblock \bibinfo{title}{{Dune Dashboard - zkSync Era}}.
\newblock
\newblock
\urldef\tempurl%
\url{https://dune.com/Marcov/zkSync}
\showURL{%
\tempurl}


\bibitem[202(2024d)]%
        {2024EthereumRoadmap}
 \bibinfo{year}{2024}\natexlab{d}.
\newblock \bibinfo{title}{{Ethereum Roadmap}}.
\newblock
\newblock
\urldef\tempurl%
\url{https://ethereum.org/en/roadmap/}
\showURL{%
\tempurl}


\bibitem[Adams et~al\mbox{.}(2021)]%
        {adams2021uniswap}
\bibfield{author}{\bibinfo{person}{Hayden Adams}, \bibinfo{person}{Noah Zinsmeister}, \bibinfo{person}{Moody Salem}, \bibinfo{person}{River Keefer}, {and} \bibinfo{person}{Dan Robinson}.} \bibinfo{year}{2021}\natexlab{}.
\newblock \bibinfo{title}{{Uniswap v3 core}}.
\newblock
\newblock


\bibitem[Angelis and Da~Silva(2019)]%
        {angelis2019blockchain}
\bibfield{author}{\bibinfo{person}{Jannis Angelis} {and} \bibinfo{person}{Elias~Ribeiro Da~Silva}.} \bibinfo{year}{2019}\natexlab{}.
\newblock \showarticletitle{Blockchain adoption: A value driver perspective}.
\newblock \bibinfo{journal}{\emph{Business Horizons}} \bibinfo{volume}{62}, \bibinfo{number}{3} (\bibinfo{year}{2019}), \bibinfo{pages}{307--314}.
\newblock


\bibitem[Anonymous(2023)]%
        {Domo@BRC}
\bibfield{author}{\bibinfo{person}{Anonymous}.} \bibinfo{year}{2023}\natexlab{}.
\newblock \showarticletitle{BRC-20 experiment}.
\newblock  (\bibinfo{year}{2023}).
\newblock
\urldef\tempurl%
\url{https://domo-2.gitbook.io/brc-20-e xperiment/}
\showURL{%
\tempurl}


\bibitem[Bertucci(2023)]%
        {bertucci2023bitcoin}
\bibfield{author}{\bibinfo{person}{Louis Bertucci}.} \bibinfo{year}{2023}\natexlab{}.
\newblock \showarticletitle{Bitcoin Ordinals: Determinants and Impact on Total Transaction Fees}.
\newblock \bibinfo{journal}{\emph{Available at SSRN 4486127}} (\bibinfo{year}{2023}).
\newblock


\bibitem[Daian et~al\mbox{.}(2020)]%
        {Daian@S&P20}
\bibfield{author}{\bibinfo{person}{Philip Daian}, \bibinfo{person}{Steven Goldfeder}, \bibinfo{person}{Tyler Kell}, \bibinfo{person}{Yunqi Li}, \bibinfo{person}{Xueyuan Zhao}, \bibinfo{person}{Iddo Bentov}, \bibinfo{person}{Lorenz Breidenbach}, {and} \bibinfo{person}{Ari Juels}.} \bibinfo{year}{2020}\natexlab{}.
\newblock \showarticletitle{Flash Boys 2.0: Frontrunning in Decentralized Exchanges, Miner Extractable Value, and Consensus Instability}. In \bibinfo{booktitle}{\emph{2020 IEEE Symposium on Security and Privacy (SP)}}.
\newblock


\bibitem[{Ethereum Foundation}(2023)]%
        {NFTs}
\bibfield{author}{\bibinfo{person}{{Ethereum Foundation}}.} \bibinfo{year}{2023}\natexlab{}.
\newblock \bibinfo{title}{{Non-fungible tokens (NFT)}}.
\newblock \bibinfo{howpublished}{\url{https://ethereum.org/en/nft}}.
\newblock
\newblock
\shownote{Accessed on April 15, 2024}.


\bibitem[Gangwal et~al\mbox{.}(2022)]%
        {gangwal2022survey}
\bibfield{author}{\bibinfo{person}{Ankit Gangwal}, \bibinfo{person}{Haripriya~Ravali Gangavalli}, {and} \bibinfo{person}{Apoorva Thirupathi}.} \bibinfo{year}{2022}\natexlab{}.
\newblock \showarticletitle{A Survey of Layer-Two Blockchain Protocols}.
\newblock  (\bibinfo{year}{2022}).
\newblock
\showeprint[arxiv]{2204.08032}~[cs.CR]


\bibitem[Leshner and Hayes(2019)]%
        {leshner2019compound}
\bibfield{author}{\bibinfo{person}{Robert Leshner} {and} \bibinfo{person}{Geoffrey Hayes}.} \bibinfo{year}{2019}\natexlab{}.
\newblock \bibinfo{title}{Compound: The money market protocol}.
\newblock
\newblock


\bibitem[{MakerDAO}(2023)]%
        {Governance@MakerDAO}
\bibfield{author}{\bibinfo{person}{{MakerDAO}}.} \bibinfo{year}{2023}\natexlab{}.
\newblock \bibinfo{title}{{Governance Module -- Maker Protocol Technical Docs}}.
\newblock \bibinfo{howpublished}{\url{https://docs.makerdao.com/smart-contract-modules/governance-module}}.
\newblock
\newblock
\shownote{Accessed on April 2, 2023}.


\bibitem[Messias et~al\mbox{.}(2023)]%
        {messias2023understanding}
\bibfield{author}{\bibinfo{person}{Johnnatan Messias}, \bibinfo{person}{Vabuk Pahari}, \bibinfo{person}{Balakrishnan Chandrasekaran}, \bibinfo{person}{Krishna~P Gummadi}, {and} \bibinfo{person}{Patrick Loiseau}.} \bibinfo{year}{2023}\natexlab{}.
\newblock \showarticletitle{Understanding blockchain governance: Analyzing decentralized voting to amend defi smart contracts}.
\newblock \bibinfo{journal}{\emph{arXiv preprint arXiv:2305.17655}} (\bibinfo{year}{2023}).
\newblock


\bibitem[Nakamoto(2008)]%
        {Nakamoto-WhitePaper2008}
\bibfield{author}{\bibinfo{person}{Satoshi Nakamoto}.} \bibinfo{year}{2008}\natexlab{}.
\newblock \bibinfo{title}{{Bitcoin: A Peer-to-Peer Electronic Cash System}}.
\newblock
\newblock


\bibitem[Perez et~al\mbox{.}(2021)]%
        {Perez@FC21}
\bibfield{author}{\bibinfo{person}{Daniel Perez}, \bibinfo{person}{Sam~M Werner}, \bibinfo{person}{Jiahua Xu}, {and} \bibinfo{person}{Benjamin Livshits}.} \bibinfo{year}{2021}\natexlab{}.
\newblock \showarticletitle{Liquidations: DeFi on a Knife-edge}. In \bibinfo{booktitle}{\emph{Financial Cryptography and Data Security}} \emph{(\bibinfo{series}{FC '21})}.
\newblock


\bibitem[Prewett et~al\mbox{.}(2020)]%
        {prewett2020blockchain}
\bibfield{author}{\bibinfo{person}{Kyleen~W Prewett}, \bibinfo{person}{Gregory~L Prescott}, {and} \bibinfo{person}{Kirk Phillips}.} \bibinfo{year}{2020}\natexlab{}.
\newblock \showarticletitle{Blockchain adoption is inevitable—Barriers and risks remain}.
\newblock \bibinfo{journal}{\emph{Journal of Corporate accounting \& finance}} \bibinfo{volume}{31}, \bibinfo{number}{2} (\bibinfo{year}{2020}), \bibinfo{pages}{21--28}.
\newblock


\bibitem[{Provenance}(2015)]%
        {Provenance-WebArticle2015}
\bibfield{author}{\bibinfo{person}{{Provenance}}.} \bibinfo{year}{2015}\natexlab{}.
\newblock \bibinfo{title}{{Blockchain: the solution for transparency in product supply chains}}.
\newblock \bibinfo{howpublished}{\url{https://www.provenance.org/whitepaper}}.
\newblock


\bibitem[Qin et~al\mbox{.}(2021)]%
        {Qin@FC21}
\bibfield{author}{\bibinfo{person}{Kaihua Qin}, \bibinfo{person}{Liyi Zhou}, \bibinfo{person}{Benjamin Livshits}, {and} \bibinfo{person}{Arthur Gervais}.} \bibinfo{year}{2021}\natexlab{}.
\newblock \showarticletitle{{Attacking the DeFi Ecosystem with Flash Loans for Fun and Profit}}. In \bibinfo{booktitle}{\emph{Financial Cryptography and Data Security}} \emph{(\bibinfo{series}{FC '21})}.
\newblock


\bibitem[{Robert Hackett}(2017)]%
        {Hackett-WebArticle2017}
\bibfield{author}{\bibinfo{person}{{Robert Hackett}}.} \bibinfo{year}{2017}\natexlab{}.
\newblock \bibinfo{title}{{Walmart and 9 Food Giants Team Up on IBM Blockchain Plans}}.
\newblock \bibinfo{howpublished}{\url{http://fortune.com/2017/08/22/walmart-blockchain-ibm-food-nestle-unilever-tyson-dole}}.
\newblock


\bibitem[Sguanci et~al\mbox{.}(2021)]%
        {sguanci2021layer}
\bibfield{author}{\bibinfo{person}{Cosimo Sguanci}, \bibinfo{person}{Roberto Spatafora}, {and} \bibinfo{person}{Andrea~Mario Vergani}.} \bibinfo{year}{2021}\natexlab{}.
\newblock \showarticletitle{Layer 2 blockchain scaling: A survey}.
\newblock \bibinfo{journal}{\emph{arXiv preprint arXiv:2107.10881}} (\bibinfo{year}{2021}).
\newblock


\bibitem[{Shivam Sharm}(2023)]%
        {BRC-20@Binance}
\bibfield{author}{\bibinfo{person}{{Shivam Sharm}}.} \bibinfo{year}{2023}\natexlab{}.
\newblock \bibinfo{title}{{BRC-20 Tokens: A Primer}}.
\newblock \bibinfo{howpublished}{\url{https://www.binance.com/en/research/analysis/brc-20-tokens-a-primer}}.
\newblock
\newblock
\shownote{Accessed on Jan 15, 2023}.


\bibitem[{Tom Blackstone}(2023)]%
        {Tom-Arbitrum@Cointelegraph}
\bibfield{author}{\bibinfo{person}{{Tom Blackstone}}.} \bibinfo{year}{2023}\natexlab{}.
\newblock \bibinfo{title}{{Arbitrum network went offline for 78 minutes because of inscriptions}}.
\newblock \bibinfo{howpublished}{\url{https://cointelegraph.com/news/arbitrum-network-goes-offline-december-15}}.
\newblock
\newblock
\shownote{Accessed on Jan 15, 2023}.


\bibitem[Wang et~al\mbox{.}(2021)]%
        {wang2021non}
\bibfield{author}{\bibinfo{person}{Qin Wang}, \bibinfo{person}{Rujia Li}, \bibinfo{person}{Qi Wang}, {and} \bibinfo{person}{Shiping Chen}.} \bibinfo{year}{2021}\natexlab{}.
\newblock \showarticletitle{Non-fungible token (NFT): Overview, evaluation, opportunities and challenges}.
\newblock \bibinfo{journal}{\emph{arXiv preprint arXiv:2105.07447}} (\bibinfo{year}{2021}).
\newblock


\bibitem[Wang and Yu(2023a)]%
        {wang2023brc}
\bibfield{author}{\bibinfo{person}{Qin Wang} {and} \bibinfo{person}{Guangsheng Yu}.} \bibinfo{year}{2023}\natexlab{a}.
\newblock \showarticletitle{BRC-20: Hope or Hype}.
\newblock \bibinfo{journal}{\emph{arXiv preprint arXiv:2310.10652}} (\bibinfo{year}{2023}).
\newblock


\bibitem[Wang and Yu(2023b)]%
        {wang2023understanding}
\bibfield{author}{\bibinfo{person}{Qin Wang} {and} \bibinfo{person}{Guangsheng Yu}.} \bibinfo{year}{2023}\natexlab{b}.
\newblock \showarticletitle{Understanding BRC-20: Hope or hype}.
\newblock \bibinfo{journal}{\emph{Available at SSRN 4590451}} (\bibinfo{year}{2023}).
\newblock


\bibitem[Werth et~al\mbox{.}(2023)]%
        {werth2023review}
\bibfield{author}{\bibinfo{person}{Jan Werth}, \bibinfo{person}{Mohammad~Hajian Berenjestanaki}, \bibinfo{person}{Hamid~R Barzegar}, \bibinfo{person}{Nabil El~Ioini}, {and} \bibinfo{person}{Claus Pahl}.} \bibinfo{year}{2023}\natexlab{}.
\newblock \showarticletitle{A Review of Blockchain Platforms Based on the Scalability, Security and Decentralization Trilemma.}
\newblock \bibinfo{journal}{\emph{ICEIS (1)}} (\bibinfo{year}{2023}), \bibinfo{pages}{146--155}.
\newblock


\bibitem[Wood(2014)]%
        {Wood@Ethereum}
\bibfield{author}{\bibinfo{person}{Gavin Wood}.} \bibinfo{year}{2014}\natexlab{}.
\newblock \bibinfo{title}{Ethereum: A secure decentralised generalised transaction ledger}.
\newblock
\newblock


\bibitem[Xu et~al\mbox{.}(2023)]%
        {xu2023sok}
\bibfield{author}{\bibinfo{person}{Jiahua Xu}, \bibinfo{person}{Krzysztof Paruch}, \bibinfo{person}{Simon Cousaert}, {and} \bibinfo{person}{Yebo Feng}.} \bibinfo{year}{2023}\natexlab{}.
\newblock \showarticletitle{Sok: Decentralized exchanges (dex) with automated market maker (amm) protocols}.
\newblock \bibinfo{journal}{\emph{Comput. Surveys}} \bibinfo{volume}{55}, \bibinfo{number}{11} (\bibinfo{year}{2023}), \bibinfo{pages}{1--50}.
\newblock


\bibitem[Yee et~al\mbox{.}(2022)]%
        {yee2022shades}
\bibfield{author}{\bibinfo{person}{Bennet Yee}, \bibinfo{person}{Dawn Song}, \bibinfo{person}{Patrick McCorry}, {and} \bibinfo{person}{Chris Buckland}.} \bibinfo{year}{2022}\natexlab{}.
\newblock \showarticletitle{Shades of Finality and Layer 2 Scaling}.
\newblock  (\bibinfo{year}{2022}).
\newblock
\showeprint[arxiv]{2201.07920}~[cs.CR]


\bibitem[Yu and Wang(2023)]%
        {yu2023bridging}
\bibfield{author}{\bibinfo{person}{Guangsheng Yu} {and} \bibinfo{person}{Qin Wang}.} \bibinfo{year}{2023}\natexlab{}.
\newblock \showarticletitle{Bridging BRC-20 to Ethereum}.
\newblock \bibinfo{journal}{\emph{arXiv preprint arXiv:2310.10065}} (\bibinfo{year}{2023}).
\newblock


\end{thebibliography}
